\documentclass[
    aps,
    prl,
    twocolumn,
    superscriptaddress,
    usenames,
    dvipsnames,
    floatfix]{revtex4-1}

\usepackage{graphicx}
\usepackage{bm}
\usepackage{commath}
\usepackage{amssymb, amsfonts, amsmath}
\usepackage{xcolor}
\usepackage{soul}
\usepackage{hyperref}

\newcommand{\lnkvidone}{https://youtu.be/nDTasAbLDCo}
\newcommand{\lnkvidtwo}{https://youtu.be/DBb4Ua0lQKc}
\newcommand{\lnkvidthree}{https://youtu.be/OqLJn9MVFNE}
\newcommand{\lnkvidfour}{https://youtu.be/Gm7J8bc3TVI}
\newcommand{\lnkvidfive}{https://youtu.be/f9EOdHH2Tv8}
\newcommand{\lnkvidsix}{https://youtu.be/FFU5vDQ33TA}

\newcommand{\total}{{\mathrm{d}}}
\newcommand{\bs}[1]{{\boldsymbol{#1}}}
\renewcommand{\tilde}[1]{{\widetilde{#1}}}

\begin{document}

\title{Transitions of turbulent superstructures in generalized Kolmogorov flow}

\author{Cristian C. Lalescu}
\affiliation{Max Planck Institute for Dynamics and Self-Organization, Am Fa\ss berg 17, 37077 G\"ottingen}
\affiliation{Max Planck Computing and Data Facility, Gie\ss enbachstra\ss e 2, 85748, Garching b. M\"unchen}

\author{Michael Wilczek}
\email{michael.wilczek@ds.mpg.de}
\affiliation{Max Planck Institute for Dynamics and Self-Organization, Am Fa\ss berg 17, 37077 G\"ottingen}
\affiliation{Institute for the Dynamics of Complex Systems, University of G\"ottingen, Friedrich-Hund-Platz 1, 37077 G\"ottingen}

\date{\today}

\begin{abstract}
Self-organized large-scale flow structures occur in a wide range of turbulent flows. Yet, their emergence, dynamics, and interplay with small-scale turbulence are not well understood. Here, we investigate such self-organized turbulent superstructures in three-dimensional turbulent Kolmogorov flow with large-scale drag. Through extensive simulations, we uncover their low-dimensional dynamics featuring transitions between several stable and meta-stable large-scale structures as a function of the damping parameter. The main dissipation mechanism for the turbulent superstructures is the generation of small-scale turbulence, whose local structure depends strongly on the large-scale flow. Our results elucidate the generic emergence and low-dimensional dynamics of large-scale flow structures in fully developed turbulence and reveal a strong coupling of large- and small-scale flow features.
\end{abstract}

\maketitle

%**********************************************************
% Introduction
%**********************************************************

Most turbulent flows, including all large-scale flows in our oceans and our atmosphere, are geometrically constrained and driven by anisotropic large-scale forces such as temperature gradients, rotation or shear. While their small scales are typically fully turbulent, the large scales often show strikingly coherent flow patterns, termed turbulent superstructures. In boundary layer flows, turbulent superstructures appear in the form of very large-scale meandering low- and high-velocity streaks \cite{marusic2010science,hellstroem2015jfm,marusic2017prf}. Convective flows, such as Rayleigh-B\'enard convection, exhibit persistent large-scale convection rolls even in the fully turbulent regime \cite{hartlep2003prl, parodi2004prl, hartlep2005jfm,vonhardenberg2008pla, emran2015jfm, stevens2018prf, pandey2018nc,green20jfm,wang2020prl}. Coherent large-scale flow states have also been observed in turbulent Taylor-Couette flow \cite{huisman2014nc,ostillamonico2015prf,vanderveen2016prf,sacco2019jfm}, turbulent plane Couette flow \cite{xia2018jfm} as well as in von-K\'arm\'an flow \cite{ravelet2004prl,ravelet08jfm,cortet2010prl}. Superstructures in fully developed turbulence therefore are a widely occurring, dynamically emergent phenomenon.

Despite their generic occurrence in a broad range of prototypical flows, the emergence and dynamics of turbulent superstructures, and in particular their interplay with small-scale turbulence, are currently not well understood. However, these aspects are crucial for developing a low-dimensional description of the large-scale dynamics of fully developed turbulent flows. Moreover, Landau pointed out in a famous remark on Kolmogorov's 1941 phenomenology \cite{landau_lifshitz_fluid1987} that the large-scale structure of a flow impacts the temporal variation of the rate of energy dissipation, potentially precluding a universal statistical theory of turbulence valid for all flows. Therefore, a characterization of the coupling of large and small scales has far-reaching implications for the assessment of universal features (or lack thereof) of small-scale turbulence \cite{chien2013jfm}.

The need to simultaneously resolve the slowly evolving large scales and the rapidly fluctuating small scales renders this problem challenging, both for experiments as well as for simulations. To address this, we here study a prototypical shear flow---a generalized three-dimensional (3D) turbulent Kolmogorov flow, which allows the investigation of turbulent superstructures without the complications imposed by boundaries. Traditional Kolmogorov flow is a simple, two-dimensional (2D) shear flow driven by a single Fourier mode, originally proposed by Kolmogorov to study the onset of turbulence on a periodic domain \cite{meshalkin1961pmm}. Indeed, the onset of turbulence \cite{thess92pfa} (for 3D see \cite{veen2016fdr})
as well as the chaotic dynamics  significantly above the onset \cite{chandler13jfm,lucas14jfm,lucas15pof}
have been studied in detail. In contrast to the simple two-dimensional setting, the direct energy cascade toward smaller scales in three dimensions generates fully developed small-scale turbulence. Computational studies of turbulent 3D Kolmogorov flow revealed pronounced spatio-temporal large-scale intermittency and inhomogeneity \cite{borue1996jfm, musacchio2014pre}, anisotropy \cite{iyer17prf}, as well as translational symmetry breaking as a function of domain size~\cite{sarris2007pof}.

Here, we generalize 3D Kolmogorov flow by including a large-scale drag term, which allows the manipulation of the range of scales on which turbulent superstructures occur. We reveal their low-dimensional dynamics in fully developed turbulence encompassing millions to billions of degrees of freedom by means of extensive simulations.
In particular, we observe transitions between several
large-scale structures as a function of the damping parameter.
Remarkably, we find that the flow is most effectively driven for strong damping, when the large-scale flow is shaped to resonate with the shear forcing.
The generation of small-scale turbulence acts as main dissipation channel for the large-scale flow. A detailed spatio-temporal analysis of the energy transfer between large and small scales reveals a strong coupling of large- and small-scale flow features even in fully turbulent flows. In particular, we find that the mean rate of energy dissipation is well correlated with the energy injection rate by the large-scale forcing with a time lag associated to the energy transfer across scales.
This establishes generalized Kolmogorov flow as a prototypical incarnation of a flow in which the large-scale flow variation impacts the mean rate of kinetic energy dissipation as conjectured by Landau \cite{landau_lifshitz_fluid1987}.

%**********************************************************
% System & Simulations
%**********************************************************

Our Kolmogorov flow is governed by the incompressible Navier-Stokes equation
\begin{equation}
    \partial_t \bs u
    +\bs u \cdot \nabla \bs u
    =
    - \nabla p
    + \nu \Delta \bs u
    + \bs f
    - \mu \tilde{\bs u},
    \label{eq:Kflow_with_drag}
\end{equation}
on a periodic $6\pi \times 2\pi \times 2\pi$ domain. Here, $\bs u$ denotes the incompressible velocity field ($\nabla \cdot \bs u=0$), and $p$ the kinematic pressure. The shear force $\bs f = A \sin(k_f y) {\bs e}_x$  forces the flow on a single mode, corresponding to the smallest spanwise wave number $k_f=1$, with $A=1/2$ (code units).
The linear damping term, controlled by the parameter $\mu$, only affects the sharp-Fourier-filtered field $\tilde{\bs u}$, which only contains spatial scales larger than the forcing scale. This permits the manipulation of large-scale flow structures through the damping parameter, while allowing for freely evolving small scales. The flow is also subject to viscous dissipation with the kinematic viscosity $\nu$.

To investigate the dynamics of turbulent superstructures, we conduct an extensive series of pseudo-spectral direct numerical simulations (DNS). We confirmed the robustness of our results for Reynolds numbers up to $Re \approx 1.8 \times 10^4$ [with $Re = UL / \nu$ based on the forcing scales $L = 2\pi/k_f$ and $U = (LA)^{1/2}$; Supplemental Material (SM)]. In the following, we focus on the intermediate Reynolds number of $Re \approx 7180$.

Important features of the flow can be assessed from the 2D dynamics in the forcing plane, which we obtain by decomposing the velocity field into its $z$-averaged part and fluctuations, $\bs u = \bs v + w \bs e_z + \bs u'$.
Here, $\bs v$ denotes the $z$-averaged, 2D velocity field in the forcing plane, and $w$ is the $z$-averaged vertical (out-of-plane) velocity component, which is expected to be generally small; $\bs u'$ denotes the 3D velocity fluctuations.
The averaged 2D dynamics follows (see also \cite{alexakis2018pr}):
\begin{equation}
        \partial_t \bs v + \bs v \cdot \nabla \bs v = 
    - \nabla \langle p \rangle_z
    + \nu \Delta \bs v + \bs f - \mu \tilde{\bs v} - \nabla \cdot \langle \bs u' \bs u' \rangle_z \, ,
    \label{eq:zaveraged_Kflow_with_drag}
\end{equation}
where $\langle \cdot \rangle_z$ denotes $z$-averaging.  Apart from interactions mediated by the 3D fluctuations through the stresses, from which only horizontal (in-plane) velocity fluctuations contribute to this equation, the vertical velocity $w$ does not contribute to the dynamics of the 2D velocity $\bs v$.

%**********************************************************
% Results
%**********************************************************

% Fig 1: different large-scale states

\begin{figure}
    \includegraphics[width=\columnwidth]{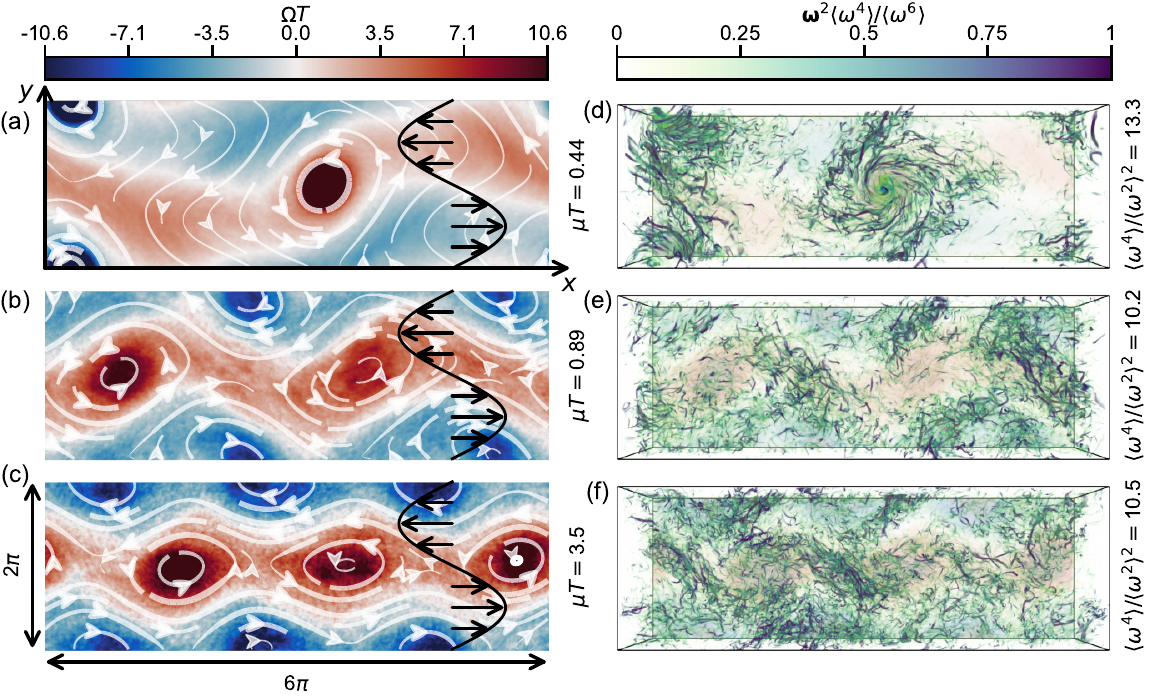}
    \caption{
        Turbulent Kolmogorov flow on an aspect ratio three domain at $Re \approx 7180$.
        (a-c):
        The system exhibits different large-scale flow patterns as a function of the damping parameter $\mu$.
        The visualization shows the $z$-averaged vorticity $\Omega$ along with streamlines of the 2D flow, averaged over a time window in which the large-scale pattern is persistent (see text for details, $T=L/U$).
        (d-f): Volume renderings of the instantaneous vorticity field show that the spatial distribution of small-scale vortex structures and the corresponding vorticity flatness depend on the large-scale flow state (Supplemental Videos
        \href{\lnkvidone}{1} and
        \href{\lnkvidtwo}{2}).
    }
    \label{fig:Kolmogorov_flow}
\end{figure}

From our simulations, we observe the emergence of large-scale vortex structures in the forcing plane, which we characterize by the vorticity of the 2D averaged velocity, $\Omega = \partial_x v_y - \partial_y v_x$. Figure \ref{fig:Kolmogorov_flow} shows a range of different large-scale states along with visualizations of the full 3D vorticity field for three different values of $\mu$.
For small $\mu$, a single large-scale vortex pair emerges, which can be perceived as a strongly inhomogeneous agglomeration of small-scale vorticity. The emergence of this single vortex pair can be understood as a long-wavelength instability of a flow which is initially proportional to the forcing ($k_x=0$) and then destabilizes, leading to a spanwise velocity on the smallest possible wave number ($k_x = k_f/3$). Interestingly, this wave number is also expected based on a linear stability analysis of laminar 2D Kolmogorov flow \cite{meshalkin1961pmm, chatterjee2020chaos}. By increasing $\mu$, this mode can be stabilized, triggering different flow patterns. For large $\mu$, the large-scale flow predominantly organizes into three vortex pairs. The small-scale turbulence also becomes more homogeneous, resulting in a reduced vorticity flatness when compared to the single-vortex-pair state, see Fig.~\ref{fig:Kolmogorov_flow}. Since the turbulent Taylor-scale Reynolds number increases from small non-zero values of $\mu$ to larger ones, this trend is not due to an increase of small-scale turbulence (SM). Overall, this suggests that different superstructures can induce different levels of small-scale intermittency. For intermediate values of $\mu$, the flow dynamically switches between the three different states (one, two, and three vortex pairs). Therefore, the flow cannot be characterized by a single, universal large-scale flow state for a fixed set of parameters in this range. The spontaneous symmetry breaking in the form of the occurrence of one to three vortex pairs is a feature of the aspect-ratio-three domain.
In domains of aspect ratio one \cite{borue1996jfm, sarris2007pof, musacchio2014pre} the mean velocity field is proportional to the forcing and therefore invariant with respect to continuous $x$ translations. Larger-aspect-ratio domains, on the other hand, allow for an even greater variety of turbulent superstructures.

% Fig 2: large-scale dynamics

\begin{figure}
    \includegraphics[width=0.99\columnwidth]{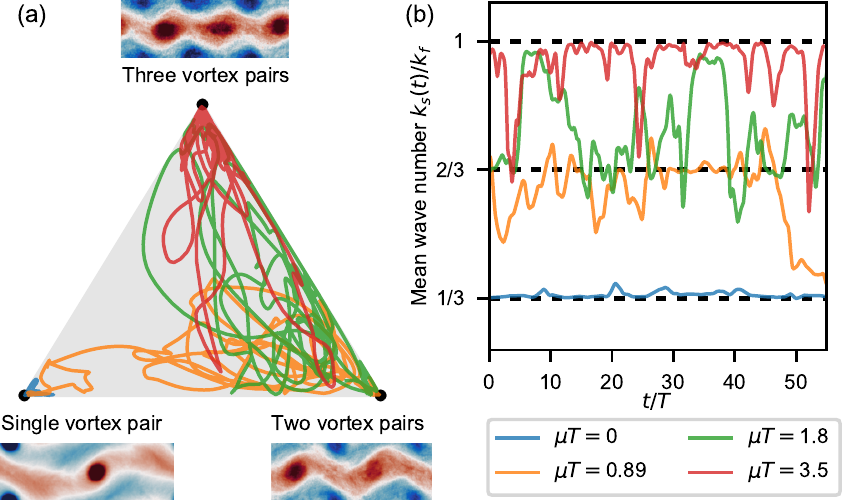}
    \caption{The large-scale flow structures exhibit a low-dimensional dynamics. Time series of the large-scale flow states characterized by their relative intensity (a) and their mean wave number (b) show that the large-scale flow states depend on the damping parameter. For $\mu=0$, the flow organizes into a single vortex pair. For intermediate damping, the flow dynamically alternates between different large-scale flow states, whereas for $\mu T=3.5$, the three-vortex-pair state is the only stable structure.
    }
    \label{fig:phase_space}
\end{figure}

To characterize the large-scale dynamics, and in particular the switching between large-scale states, we compute the stream function $\psi$, given by $\Delta \psi =  -\Omega$. From its Fourier transform $\hat \psi(k_x,k_y)$, we compute the relative intensity of the three dominant streamwise Fourier modes through $I_{j} = |\hat \psi(j k_f/3, 0)|^2/(|\hat \psi(k_f/3, 0)|^2 + |\hat \psi(2k_f/3, 0)|^2 + |\hat \psi(k_f, 0)|^2)$ with $j \in \lbrace 1,2,3 \rbrace$.
Since $I_1+I_2+I_3=1$, the large-scale dynamics takes place in a triangle, whose corners correspond to the pure one, two and three vortex-pair states.
By computing a mean streamwise wave number $k_s = k_f (I_1 + 2  I_2 +  3 I_3)/3$ and averaging the solution over intervals of time where $k_s(t)$ is approximately constant, we can identify the well-defined large-scale states shown in Fig.~\ref{fig:Kolmogorov_flow}.
Figure \ref{fig:phase_space} shows the large-scale dynamics in the intensity plane along with the mean streamwise wave number of the large-scale structures as a function of time.
For low values of the damping parameter, the large-scale flow organizes into a single-vortex pair, which from time to time destabilizes due to strong, self-induced spanwise flow (see streamlines in Fig.~\ref{fig:Kolmogorov_flow}) and then re-emerges (Supplemental Video \href{\lnkvidthree}{3}). For intermediate damping $0.67 \leq \mu T \leq 1.8$, the large-scale flow dynamically switches between the meta-stable one, two and three vortex-pair states (Supplemental Videos \href{\lnkvidfour}{4} and \href{\lnkvidfive}{5}; Supplemental Figures 3, 4 and 5).
For large values of $\mu T \gg 1.8$, we find that only the three-vortex-pair state remains accessible to the dynamics, with bursts destroying the pattern only for it to be reformed with a possible streamwise offset (Supplemental Video \href{\lnkvidsix}{6}).

% Fig 3: energy & energy transfer between large and small scales

\begin{figure}
    \includegraphics[width=\columnwidth]{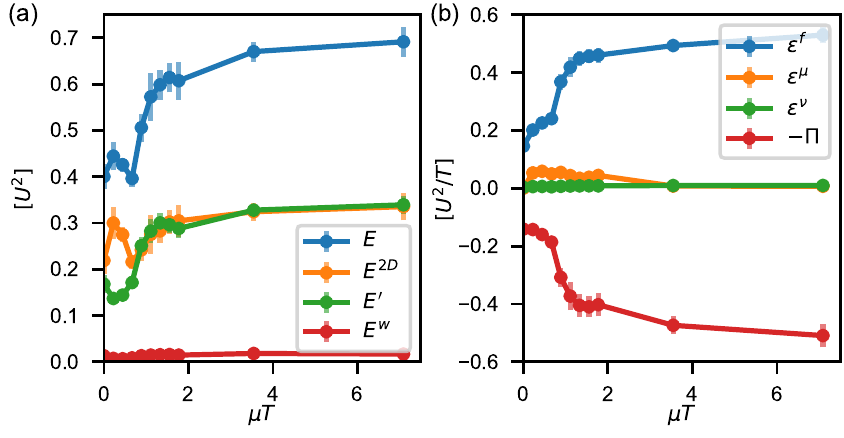}
    \caption{
        Energy budget of turbulent Kolmogorov flow.
        (a): Different contributions to the mean total energy.
        The energy contained in the 2D flow is comparable to the energy in the 3D turbulent fluctuations.
        %Remarkably, the total energy increases as a function of the damping parameter.
        (b): Energy budget of the 2D flow. The energy input into the 2D flow $\varepsilon^f$ increases with increased damping.
        The energy injection rate is almost completely balanced by the transfer of energy $\Pi$ from the 2D component to the 3D fluctuations.
        Dissipation through large-scale damping and viscous diffusion are comparably negligible.
        Error bars span one standard deviation computed with respect to temporal variations.
    }
    \label{fig:energetics}
\end{figure}

To uncover the role of small-scale turbulence for the large-scale dynamics, we investigate the energy budget of the flow.
The total energy $E(t)=\tfrac{1}{2}\left\langle \bs u \cdot \bs u \right\rangle_{xyz}$ is a sum of the contributions from the $z$-averaged 2D velocity field $E^{2D}(t)=\tfrac{1}{2}\left\langle \bs v \cdot \bs v \right\rangle_{xy}$, from the $z$-averaged vertical velocity $E^{w}(t)=\tfrac{1}{2}\left\langle w^2 \right\rangle_{xy}$, and from the remaining 3D fluctuations $E'(t) = \tfrac{1}{2}\left\langle \bs u' \cdot \bs u' \right\rangle_{xyz}$.
The energy of the 2D flow evolves according to
\begin{equation}
        \frac{\total}{\total t} E^{2D} =
        \varepsilon^{f}
        - \varepsilon^{\mu}
        - \varepsilon^{\nu}
        - \Pi  \, .
        \label{eq:energy_balance}
\end{equation}
Here, $\varepsilon^f(t) = \left\langle \bs v \cdot \bs f \right\rangle_{xy}$ denotes the energy input into the flow by the shear forcing, $\varepsilon^\mu(t) = \mu\left\langle \tilde{\bs v} \cdot \tilde{\bs v} \right\rangle_{xy}$ is the dissipation by large-scale damping, and $\varepsilon^\nu(t) = \nu\left\langle\nabla \bs v:\nabla \bs v \right\rangle_{xy}$ the viscous dissipation.
The term $\Pi(t) = -\langle \nabla \bs v : \left\langle \bs u' \bs u' \right \rangle_z \rangle_{xy}$ is the work performed by the 2D flow on the 3D stresses and characterizes the energy transfer between the 2D flow and the 3D turbulent fluctuations \cite{eyink2005pd}.
Figure \ref{fig:energetics}(a) shows the time-averaged contributions to the kinetic energy as a function of the damping parameter $\mu$. The total energy  splits up into comparable contributions from the 2D energy and the energy in 3D turbulent fluctuations, with a negligible amount of energy contained in the average vertical velocity.

Remarkably, the total energy in the flow increases as we increase $\mu$. At first, this appears counterintuitive given that we increase the strength of a dissipative term. We can explain this effect through the individual terms in the 2D budget which are shown in Fig.~\ref{fig:energetics}(b). The first important observation is that the energy input into the flow $\varepsilon^f$ increases with the damping parameter. Because $\varepsilon^f(t) = \left\langle \bs v \cdot \bs f \right\rangle_{xy}$, the energy input is maximal if $\bs v \parallel \bs f$, i.e.~if the 2D flow is a simple shear flow in streamwise direction. Out of the three large-scale states, the three-vortex-pair state is closest to a simple shear flow, as seen by comparing the streamlines in Fig.~\ref{fig:Kolmogorov_flow} with the forcing. Tuning the flow to this state by increasing the damping therefore maximizes the energy input. 

% Fig 4: where does energy transfer happen?
    \begin{figure}
        \includegraphics[width=\columnwidth]{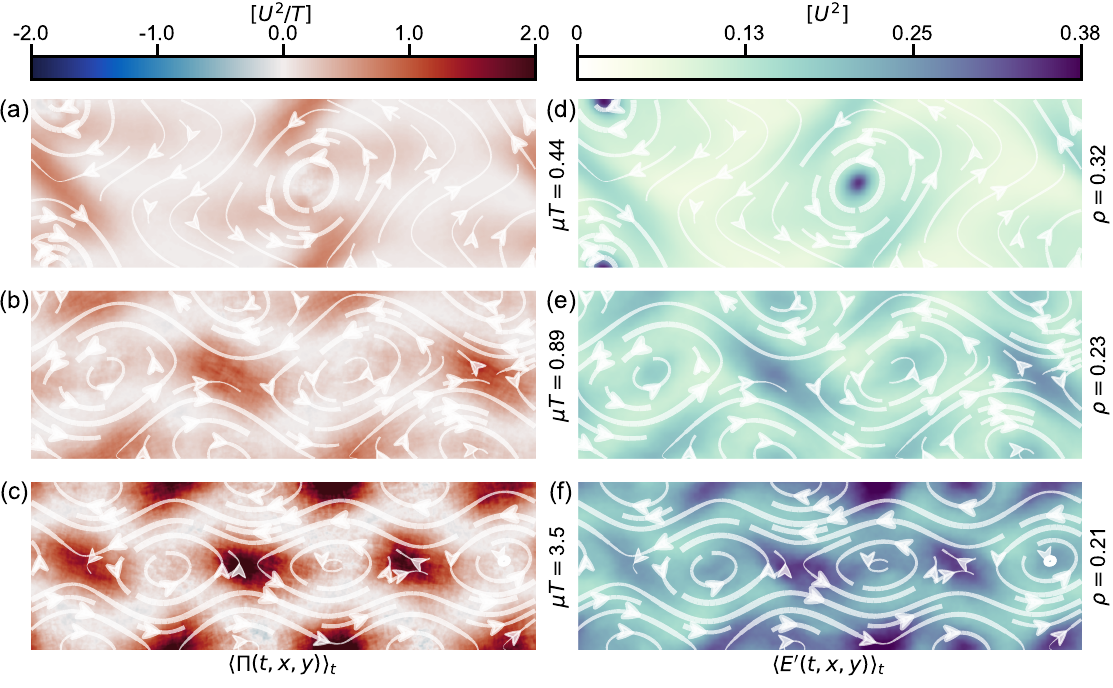}
        \caption{%
            The energy transfer and turbulent kinetic energy are strongly coupled to the large-scale flow (cf.~Fig.~\ref{fig:Kolmogorov_flow}).
            (a-c): time-averaged energy transfer term.
            Energy is predominantly transferred to 3D fluctuations at saddle points of the averaged 2D flow (indicated by streamlines).
            (d-f): time- and $z$-averaged energy of 3D fluctuations.
            The energy of the fluctuations shows strong similarities to the energy transfer term.
            The relative standard deviation $\rho$ of the fluctuation energy as a measure of inhomogeneity depends on the large-scale state.
            }
        \label{fig:spatial_structure_transfers}
    \end{figure}

Figure \ref{fig:energetics}(b) also shows that the energy input into the 2D flow is almost completely balanced by the energy transfer from the 2D flow to the 3D fluctuations. The generation of small-scale turbulence therefore is the main dissipation channel for the 2D flow. The dissipation by the large-scale damping is small for the entire parameter range, showing that its primary effect is that of shaping the large-scale flow. Because the 2D flow is predominantly large-scale, also viscous dissipation is negligible.
We have also tested the robustness of these results across Reynolds numbers (Supplemental Figure 6).

To characterize the interplay of the large-scale flow and small-scale turbulence in more detail, we analyze how the inhomogeneity of the mean flow affects the spatial structure of the local energy transfer term $\Pi(t, x, y) = - \nabla \bs v : \langle \bs u' \bs u' \rangle_z$, see Fig.~\ref{fig:spatial_structure_transfers}. The explicit argument $(t,x,y)$ indicates the coordinates over which no averaging was performed.
    The time-averaged energy transfer term is predominantly positive throughout the plane, as expected, and the large-scale state is clearly reflected in its spatial structure, with maxima close to saddle points of the mean flow.
    As this term injects energy into the 3D fluctuations, it leaves a footprint in the distribution of fluctuation energy $E'(t, x, y) = \tfrac{1}{2}\langle \bs u' \cdot \bs u' \rangle_z$.
    Further analysis confirms that, for all three states and across all values of $\mu$, correlation coefficients of at least 60\% and up to 90\% are reached for the two quantities. We quantify the inhomogeneity of the 3D fluctuations by the relative standard deviation of the energy
$\rho = \big\langle \left(\langle E' \rangle_t - \langle E' \rangle_{txy}\right)^2\big\rangle_{xy}^{1/2} / \langle E' \rangle_{txy}$
(time average taken over time spans with the same large-scale structure, computed from $E'(t, x, y)$). The values of $\rho$ are close for the two-vortex-pair and three-vortex-pair states, but significantly higher for the one-vortex-pair state. This is consistent with the vorticity flatness values given above, further highlighting the connection between the large-scale state and small-scale statistics (Supplemental Figures 3-5 and 7).

    We now turn to the temporal evolution of the energy injection, transfer and dissipation rates.
    Complementing the energy balance of $\bs{v}$, the energy budget of the spatially averaged 3D fluctuations takes the form
    \begin{equation}
            \frac{\total}{\total t} E' =
        \Pi - \xi^\nu - \xi^w \, ,
    \end{equation}
    where $\xi^\nu(t) = \nu \langle \nabla \bs u' : \nabla \bs u' \rangle_{xyz}$ denotes viscous dissipation of the 3D fluctuations and $\xi^w(t) = \left\langle \langle u'_z \bs{u}' \rangle_z \cdot \nabla w \right\rangle_{xy}$ denotes the energy transfer to the mean vertical flow. We find $\xi^w$ to be two orders of magnitude smaller than $\Pi$ and $\xi^\nu$, thus we only focus on the interplay of $\varepsilon^f$, $\Pi$ and $\xi^\nu$ in the following.

\begin{figure}[t]
        \includegraphics[width=\columnwidth]{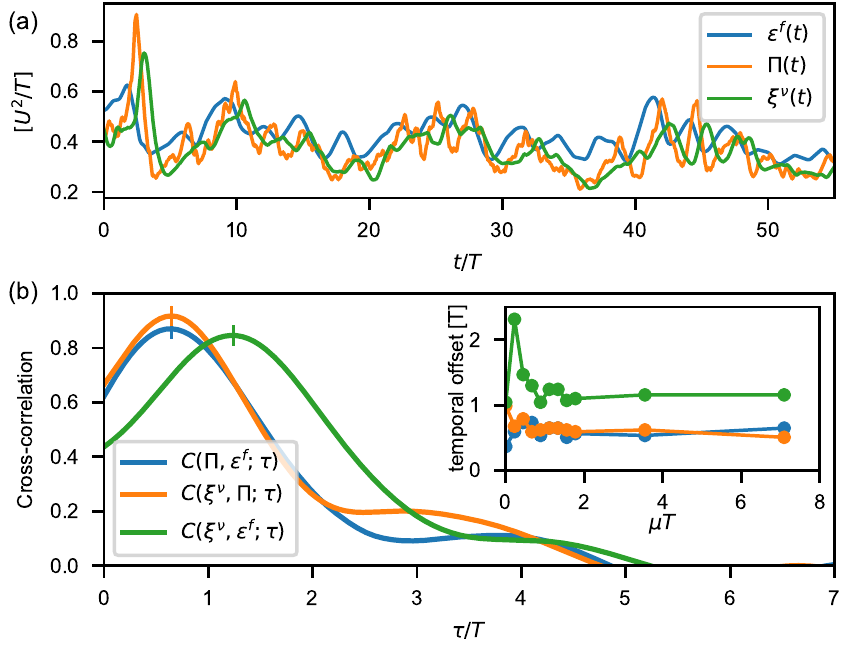}
        \caption{%
	    Cross-correlations between large-scale forcing, energy transfer, and small-scale dissipation.
        Temporal evolution of energy budget terms (top) and their cross-correlations (bottom) for $Re \approx 7180$ and $\mu T = 1.1$.
        Cross-correlations are computed to quantify the temporal offsets between the different signals (measured as the location of the maximum correlation).
        The inset shows the temporal offsets for different values of $\mu$. %(also at $Re = 7.18\times 10^3$).
            }
        \label{fig:temporal_correlations}
    \end{figure}

The temporal evolution of the self-organized large-scale flow allows us to investigate the delays between large- and small-scale flow dynamics, which are otherwise only accessible through external temporal modulation \cite{lohse2000pre,heydt2003pre}. Figure~\ref{fig:temporal_correlations} shows the time traces of the energy injection into the large-scale flow, the energy transfer term, and the viscous dissipation of the 3D fluctuations for a representative $\mu T = 1.1$ at $Re \approx 7180$ along with the normalized cross-correlations. Strong oscillations are evident, as well as a fairly strong correlation between the different time traces.

% Fig 5: how much time does the energy spend in each component?

Cross-correlations between the energy transfer term and the large-scale energy injection, $C(\Pi, \varepsilon^f;\tau)$, as well as the cross-correlations between the dissipation of 3D fluctuations and the energy transfer term, $C(\xi^\nu, \Pi;\tau)$, reach values of about 90\% for a temporal offset of approximately $0.6T$;
$\xi^\nu(t)$ is also strongly correlated with $\varepsilon^f(t)$, with a temporal offset of approximately $1.2T$.
The inset of Fig.~\ref{fig:temporal_correlations} shows the temporal offsets between the different signals as they change with the damping parameter $\mu$.
For most values of $\mu$ the offset between $\varepsilon^f(t)$ and $\xi^\nu(t)$ approximately equals the sum of the other two offsets. This can be explained by the fact that energy is transferred from the forcing scale to the dissipation scale by a direct cascade, in which the transfer between the 2D flow and the 3D fluctuations is an intermediate step. Remarkably, the two offsets are very similar in size.
These observations do not hold for very small values of $\mu$, where we also find significantly lower peak correlations between the injection rate and the dissipation rate (ca. 40-50\% rather than 80-90\%).

%**********************************************************
%Conclusions
%**********************************************************

Overall, our results draw the following picture: turbulent superstructures in Kolmogorov flow consist of accumulations of small-scale vortices, which emerge through an instability of the largest streamwise scales accessible to the dynamics. These scales can be controlled by large-scale damping. As a function of the damping parameter, we find a rich transition scenario between several large-scale states with single, two and three vortex pairs.
For intermediate damping, the system dynamically switches between various superstructures, illustrating that the system parameters do not uniquely determine the large-scale flow state. %Remarkably, we find that the flow is most effectively driven for strong damping, when the large-scale flow is shaped to resonate with the shear forcing.
The generation of 3D turbulent fluctuations is the main dissipation channel for the large-scale flow, dominating over viscous dissipation and dissipation by damping. The 2D mean flow and the 3D fluctuations are strongly coupled through the energy transfer: the topology of the 2D mean flow determines the spatial distribution of the turbulent energy transfer rate, leading to inhomogeneities in the kinetic energy of the 3D fluctuations. Additional analysis shows that these inhomogeneities likely persists even at much higher Reynolds numbers.
Generalized Kolmogorov flow is therefore one striking example of a flow in which the rates of energy transfer and ultimately dissipation vary considerably in response to the self-organization of large-scale superstructures, which has implications for the spatio-temporal structure of small-scale turbulence. Since a coupling of large- and small-scale flow features has been widely observed (see e.g.~\cite{marusic2010science,buxton2014pof,marusic2017prf,berghout2021jfm}), the scenario outlined by Landau is presumably relevant for a large class of flows.
Furthermore, our results suggest the possibility of a low-dimensional description of large-scale superstructures in fully developed turbulent flows. Since the presented mechanisms governing their emergence and dynamics are quite generic, we expect our results to be relevant for a broad range of flows including geophysical and astrophysical flows.

\acknowledgments
\section{acknowledgements}
We thank Lukas Bentkamp, Maurizio Carbone, Gerrit Green, and Johannes Zierenberg for helpful comments on the manuscript.
This work is supported by the Priority Programme SPP 1881 Turbulent Superstructures of the Deutsche Forschungsgemeinschaft. The authors gratefully acknowledge the Gauss Centre for Supercomputing e.V. for funding this project by providing computing time on the GCS Supercomputer SuperMUC at Leibniz Supercomputing Centre. Computational resources from the Max Planck Computing and Data Facility and support by the Max Planck Society are gratefully acknowledged.

\onecolumngrid
\appendix
\newpage
\renewcommand{\figurename}{Supplemental Figure}
\renewcommand{\thefigure}{S\arabic{figure}}
\renewcommand{\tablename}{Supplemental Table}
\setcounter{figure}{0}
\section{Supplemental Material}

\section{Overview of direct numerical simulations parameters}

% NOTE: these numbers vary slightly depending on which subset of iterations is used to compute statistics.

\begin{table}[h]
    \begin{tabular}{c|c|c|c|c|c|c|c|c}
        $N$ &   $Re$  & $R_\lambda'$ &  $\mu T$ &  $t_1 / T$ &  $T/\tau_\eta$ &  $L/ \eta$ &  CFL &  $k_M \eta$ \\
        \hline
        256 &   2850  & 56.8 &   0.00   &  215     &  20            &  239.5     &  0.5 &  2.7 \\
        256 &   2850  & 46.6 &   0.22   &  215     &  20            &  241.2     &  0.4 &  2.7 \\
        256 &   2850  & 47.5 &   0.44   &  213     &  22            &  248.3     &  0.4 &  2.6 \\
        256 &   2850  & 53.9 &   0.67   &  216     &  26            &  272.7     &  0.4 &  2.4 \\
        256 &   2850  & 56.7 &   0.89   &  215     &  29            &  287.1     &  0.4 &  2.2 \\
        256 &   2850  & 60.3 &   1.11   &  215     &  33            &  305.3     &  0.4 &  2.1 \\
        256 &   2850  & 59.3 &   1.33   &  216     &  34            &  309.1     &  0.5 &  2.1 \\
        256 &   2850  & 60.1 &   1.55   &  213     &  35            &  313.7     &  0.4 &  2.1 \\
        256 &   2850  & 59.4 &   1.77   &  213     &  35            &  317.7     &  0.5 &  2.0 \\
        256 &   2850  & 61.4 &   3.54   &  170     &  37            &  326.4     &  0.5 &  2.0 \\
        256 &   2850  & 61.8 &   7.09   &  166     &  38            &  328.3     &  0.5 &  2.0 \\
        256 &   2850  & 61.3 &  10.63   &  102     &  38            &  328.0     &  0.5 &  2.0 \\
        256 &   2850  & 61.9 &  14.18   &  102     &  38            &  327.7     &  0.5 &  2.0 \\
        256 &   2850  & 61.7 &  17.72   &  101     &  38            &  327.4     &  0.5 &  2.0 \\
        256 &   2850  & 62.4 &  24.81   &  102     &  38            &  328.0     &  0.5 &  2.0
    \end{tabular}
    \hfil
    \begin{tabular}{c|c|c|c|c|c|c|c|c}
        $N$  &  $Re$  & $R_\lambda'$ &   $\mu T$ &  $t_1 / T$ & $T/\tau_\eta$ &    $L/ \eta$ &    CFL &    $k_M \eta$ \\
        \hline
         512 &   7180 &   96.7  &  0.00    &  69.5      & 32            &   481.4  &  0.3  &  2.7 \\
         512 &   7180 &   77.9  &  0.22    &  67.7      & 33            &   484.1  &  0.4  &  2.7 \\
         512 &   7180 &   77.3  &  0.44    &  57.3      & 35            &   498.3  &  0.3  &  2.6 \\
         512 &   7180 &   86.0  &  0.67    &  59.1      & 37            &   515.6  &  0.3  &  2.5 \\
         512 &   7180 &   97.9  &  0.89    &  61.8      & 48            &   584.3  &  0.3  &  2.2 \\
         512 &   7180 &  100    &  1.11    &  57.7      & 52            &   612.8  &  0.4  &  2.1 \\
         512 &   7180 &  102    &  1.33    &  50.5      & 54            &   624.5  &  0.3  &  2.1 \\
         512 &   7180 &  100    &  1.55    &  50.5      & 55            &   627.4  &  0.4  &  2.1 \\
         512 &   7180 &   98.8  &  1.77    &  55.9      & 54            &   623.7  &  0.3  &  2.1 \\
         512 &   7180 &  104    &  3.54    &  54.6      & 51            &   607.2  &  0.4  &  2.1 \\
         512 &   7180 &  104    &  7.09    &   7.67     & 59            &   650.3  &  0.3  &  2.0 \\
         \hline
        1024 &  18100 &  156    &  0.00    &  10.4      & 49            &   942.1  &  0.3  &  2.7 \\
        1024 &  18100 &  116    &  0.22    &   7.90     & 53            &   975.3  &  0.4  &  2.6 \\
        1024 &  18100 &  124    &  0.44    &  10.6      & 57            &  1012.8  &  0.5  &  2.5 \\
        1024 &  18100 &  152    &  0.67    &  11.1      & 76            &  1171.3  &  0.4  &  2.2 \\
        1024 &  18100 &  155    &  0.89    &  11.3      & 76            &  1176.7  &  0.4  &  2.2 \\
        1024 &  18100 &  168    &  1.11    &  10.4      & 83            &  1226.6  &  0.4  &  2.1 \\
        1024 &  18100 &  165    &  1.33    &  10.4      & 86            &  1248.6  &  0.4  &  2.1 \\
        1024 &  18100 &  168    &  1.55    &  10.0      & 85            &  1241.7  &  0.3  &  2.1 \\
        1024 &  18100 &  172    &  1.77    &   8.46     & 88            &  1258.7  &  0.4  &  2.0 \\
        1024 &  18100 &  169    &  3.54    &   8.24     & 94            &  1304.0  &  0.4  &  2.0
    \end{tabular}
    \caption{
        Summary of direct numerical simulation series.
        Simulations were run on real-space grids with $3N \times N \times N$ points spanning the 3D periodic box $6\pi \times 2\pi \times 2\pi$ between times $t_0 = 0$ and $t_1$.
        The forcing amplitude $A$ and wave number $k_f$ are used to define large-scale units $L=2\pi/k_f$, $T = (L / A)^{1/2}$, $U = L / T$ and the Reynolds number $Re = U L / \nu$. To complement the Reynolds number based on the forcing scales, we also compute the Taylor-scale Reynolds number $R_\lambda' = 2 \langle E' \rangle_t \sqrt{5 / (3 \nu \langle \varepsilon \rangle_t)}$ using the time-averaged energy of the velocity fluctuations and the time-averaged total viscous dissipation rate.
        The time step is determined by the Courant-Friedrichs-Lewy condition, and CFL refers to the Courant number.
        The small-scale resolution of the simulations is characterized through the product of the largest resolved wave number $k_M$ and the Kolmogorov length scale $\eta$; $\tau_{\eta}$ denotes the Kolmogorov time scale.
    }
    \label{tab:simulations}
\end{table}

\newpage
\section{Supplemental videos 1 and 2 --- evolution of the 3D fluctuations}
To illustrate the structure of the 3D turbulence in
generalized Kolmogorov flow and its dependence on the large-scale flow state
(see also Fig.~1 from the main text), we provide  Supplemental Videos
\href{\lnkvidone}{1} and
\href{\lnkvidtwo}{2}
for $Re \approx 4180$. Dedicated direct numerical simulations were used to generate the required 4D datasets, with $N=256$ and $k_M \eta$ values of 2 and 1.9 for $\mu T = 0 $ and $\mu T = 3.5$, respectively.

\begin{figure}[h]
    \begin{tabular}{cc}
        \includegraphics[width=0.45\columnwidth]{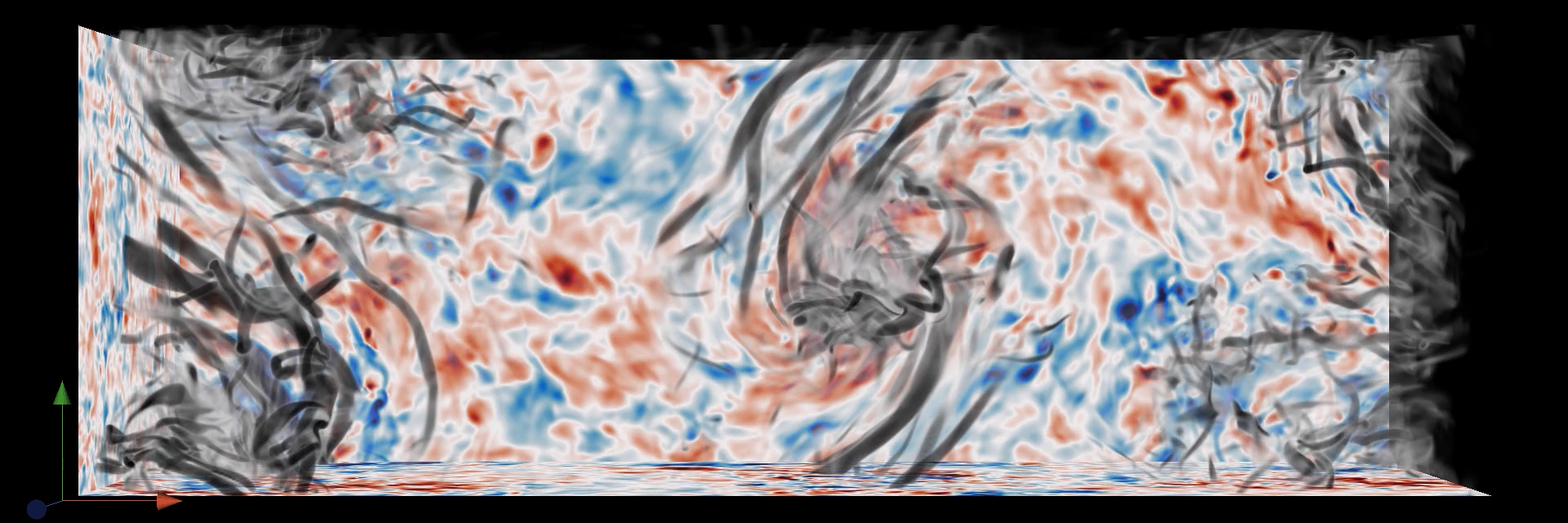}
        &
        \includegraphics[width=0.45\columnwidth]{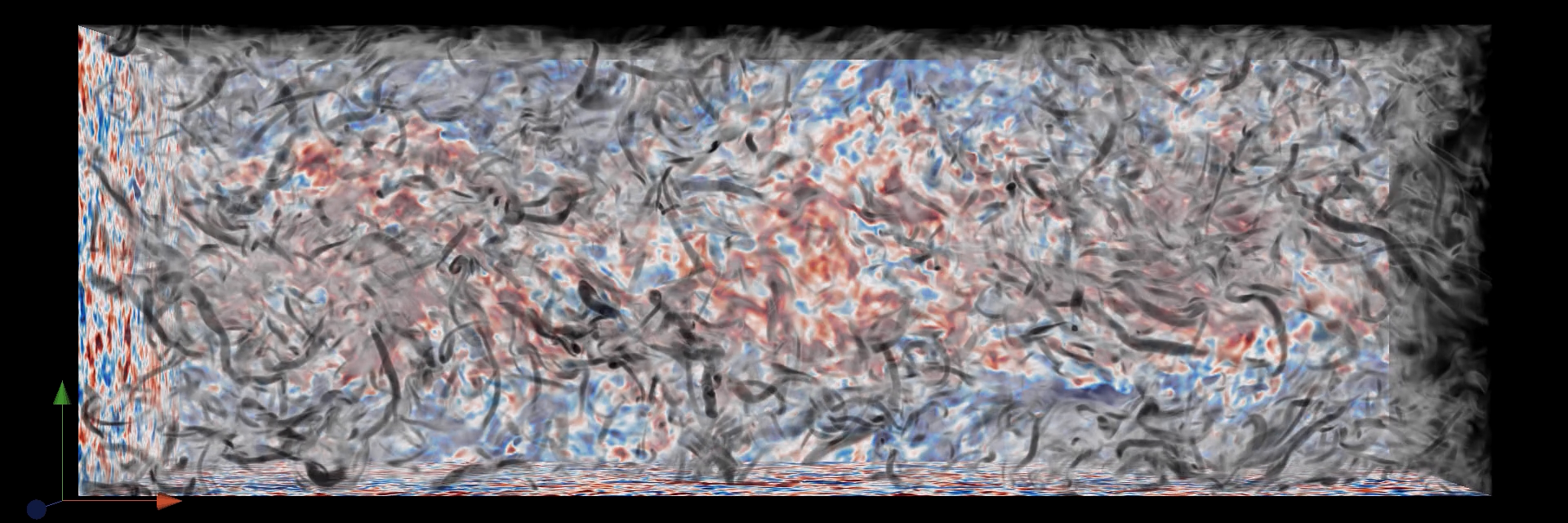}
    \end{tabular}
    \caption{%
        Evolution of the 3D fluctuations for generalized Kolmogorov flow:
        $\mu T = 0$ (left panel, \href{\lnkvidone}{Supplemental Video 1}) and $\mu T =
        3.5$ (right panel, \href{\lnkvidtwo}{Supplemental Video 2}).
        Color maps are comparable to those used for Fig.~1 from the main text.}
    \label{fig:small_scale_evolution}
\end{figure}

The videos show intense vortex filaments, visualized through volume renderings of the vorticity field $\bs \omega = \nabla \times \bs u$, see also Supplemental Fig.~\ref{fig:small_scale_evolution}. In particular, the Supplemental Figure and the videos show:
\begin{itemize}
    \item
        the $z$-averaged $z$ component of the vorticity ($xy$ plane, ``back'');
    \item
        the $x$-averaged $x$ component of the vorticity ($yz$ plane, ``left'');
    \item
        the $y$-averaged $y$ component of the vorticity ($zx$ plane, ``bottom''); the colormap corresponds to the one in Fig.~1(a)-(c) from the main text;
    \item
        the volume-rendered vorticity magnitude (colormap is the black-and-white version of the colormap used for Fig.~1(d)-(f) from the main text);
    \item
        the $xyz$ coordinate system (bottom left corner), with $x,y,z$ components colored red, green and blue, respectively.
\end{itemize}
The visualizations were generated with VTK \cite{schroeder2006book}.

\newpage

\section{Supplemental videos 3--6 --- evolution of the mean 2D flow}
\begin{figure}[h]
    \begin{tabular}{cc}
        \includegraphics[width=0.48\columnwidth]{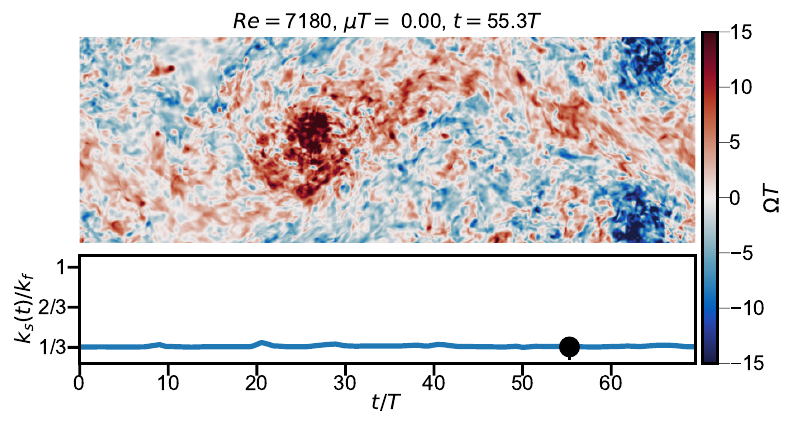}
        &
        \includegraphics[width=0.48\columnwidth]{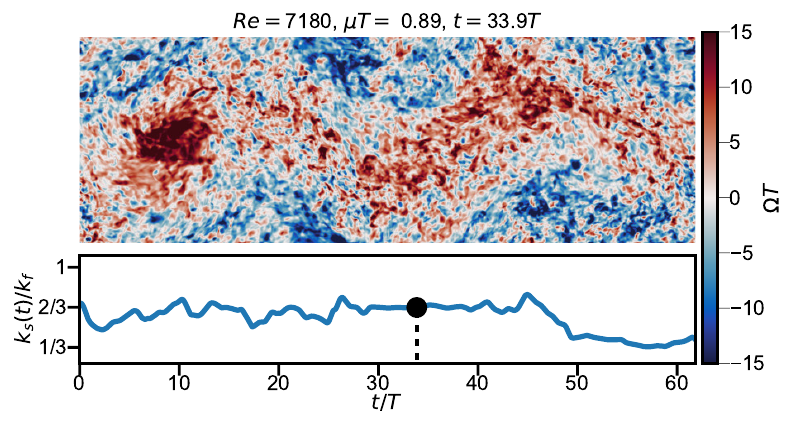} \\
        \includegraphics[width=0.48\columnwidth]{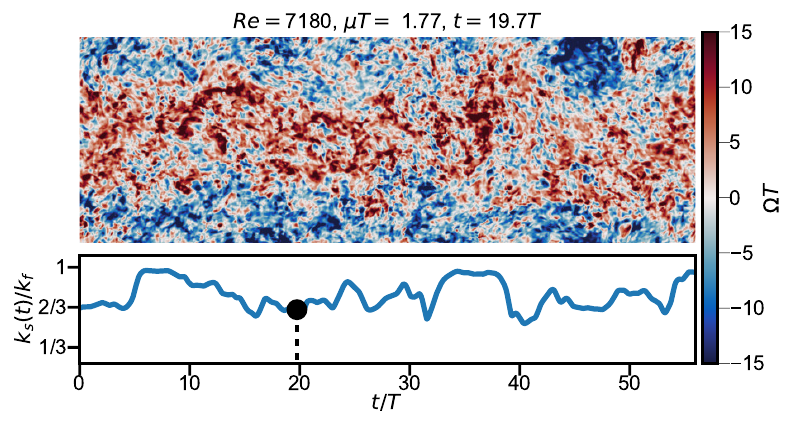}
        &
        \includegraphics[width=0.48\columnwidth]{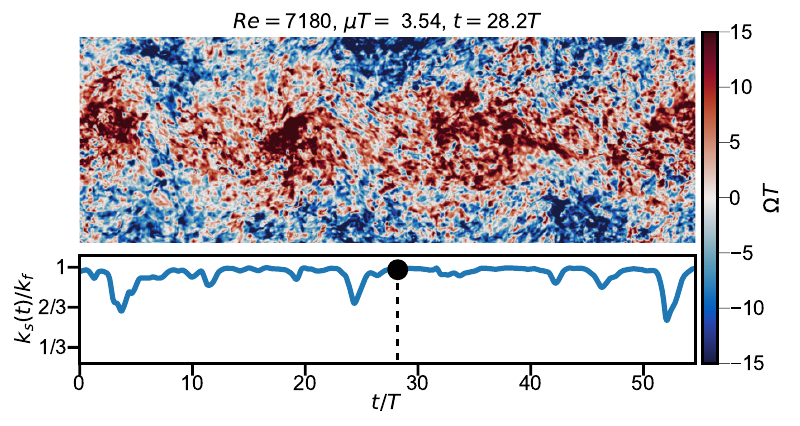}
    \end{tabular}
    \caption{
        Representative snapshots of the mean $z$ component of the vorticity $\Omega$ from Supplemental Videos
        \href{\lnkvidthree}{3},
        \href{\lnkvidfour}{4},
        \href{\lnkvidfive}{5} and
        \href{\lnkvidsix}{6}. The same four datasets used for Fig.~2 from the main text are used here.
        }
    \label{fig:Oz}
\end{figure}

The dynamics of the 2D flow $\bs v$ can be illustrated with animations of the $z$-averaged $z$ component of the vorticity $\Omega$, which we describe here.
Supplemental Fig.~\ref{fig:Oz} shows snapshots of the corresponding Supplemental Videos 3--6.

Supplemental Video \href{\lnkvidthree}{3} shows how the vortex pair in the $\mu T = 0$ case may become disorganized but always reforms.
Supplemental Videos \href{\lnkvidfour}{4}
and \href{\lnkvidfive}{5} show the large-scale dynamics for intermediate values of $\mu$ with switches between several large-scale states.
Supplemental Video \href{\lnkvidsix}{6} shows that for the comparatively large $\mu T = 3.5$ the system only visits the three-vortex-pair state.

\section{Large-scale flow states for various Reynolds numbers}

\begin{figure}
    \includegraphics[width=\columnwidth]{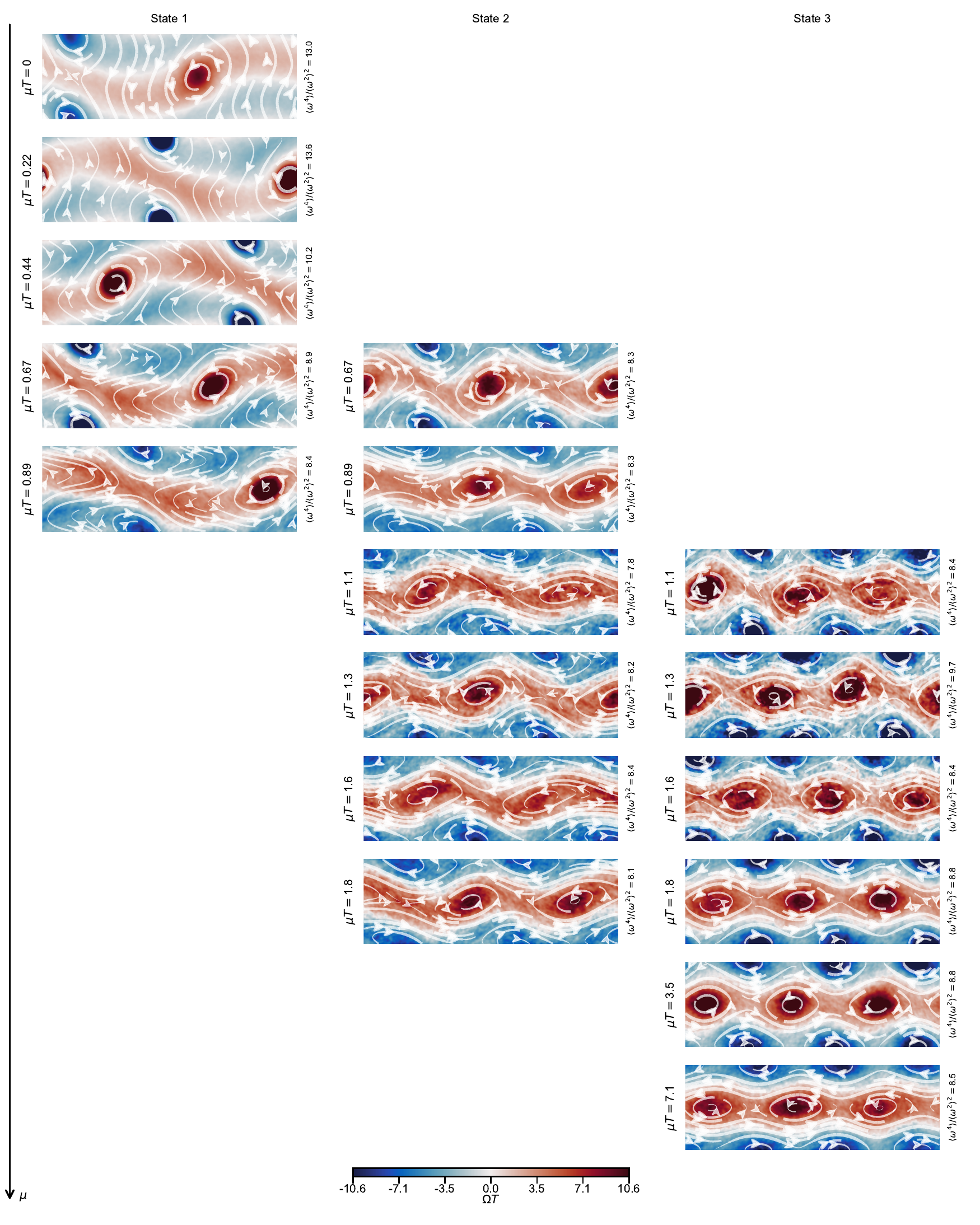}
    \caption{%
        Approximate large-scale states for $Re \approx 2850$ and different values of
        $\mu$.}
    \label{fig:flow_states_N0256}
\end{figure}

\begin{figure}
    \includegraphics[width=\columnwidth]{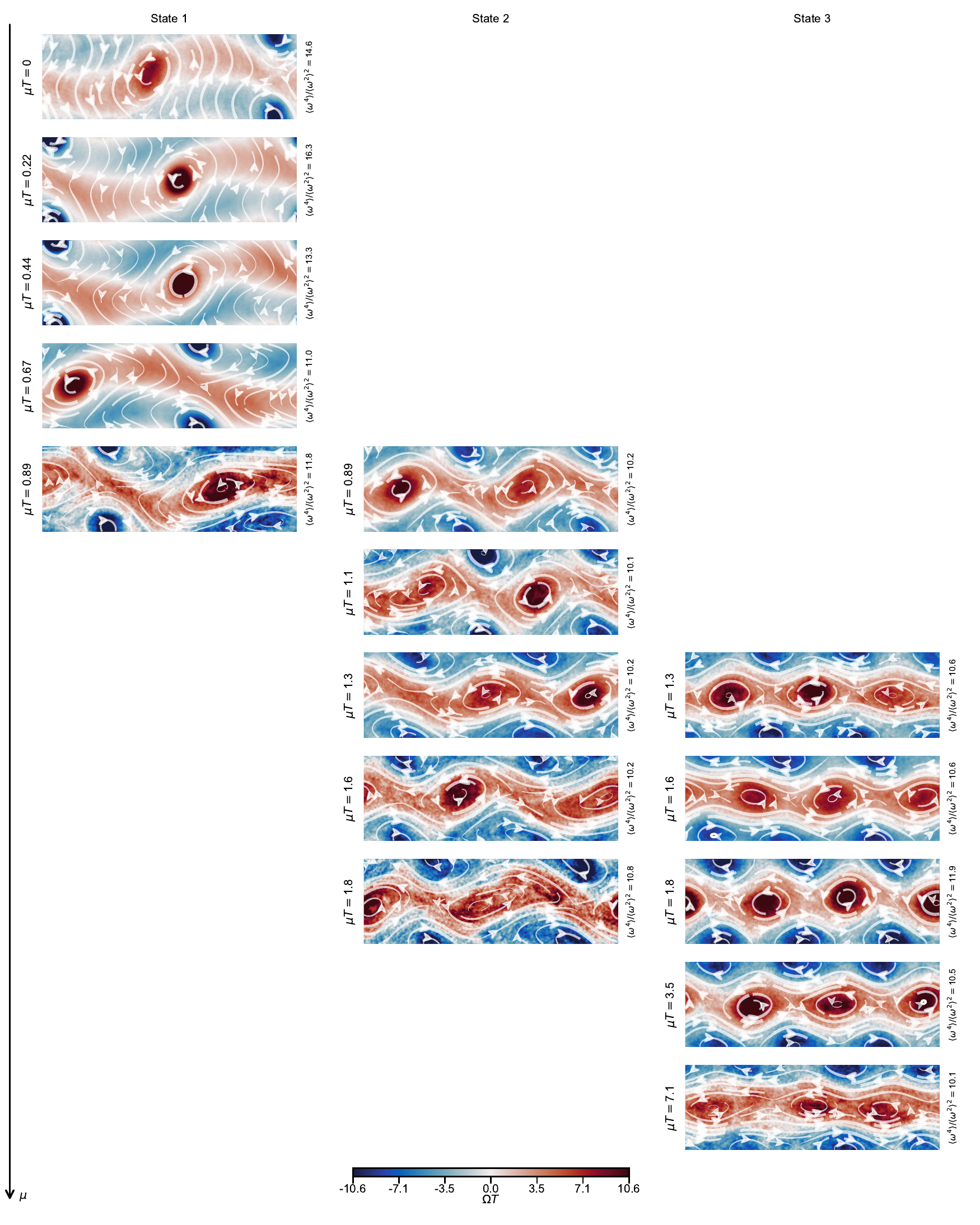}
    \caption{%
        Approximate large-scale states for $Re \approx 7180$ and different values of
        $\mu$.}
    \label{fig:flow_states_N0512}
\end{figure}

\begin{figure}
    \includegraphics[width=\columnwidth]{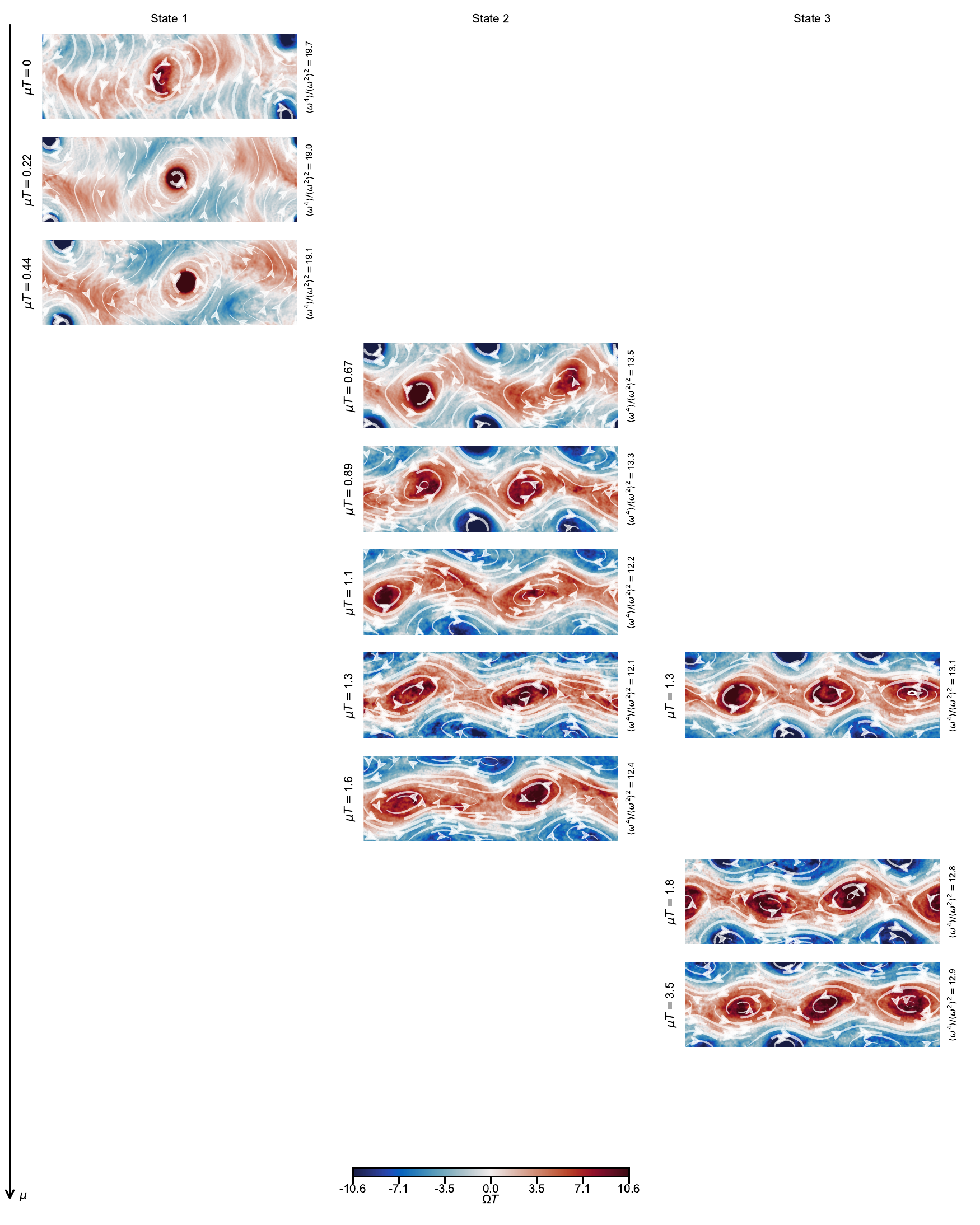}
    \caption{%
        Approximate large-scale states for $Re \approx 18100$ and different values of
        $\mu$.}
    \label{fig:flow_states_N1024}
\end{figure}

Supplemental Figs. \ref{fig:flow_states_N0256}, \ref{fig:flow_states_N0512} and \ref{fig:flow_states_N1024} provide an overview of the large-scale states observed for generalized Kolmogorov flow. Like Fig.~1(a)-(c) from the main text, the visualizations show the $z$-averaged out-of-plane vorticity along with streamlines of the 2D flow, averaged over a time window in which the large-scale pattern is persistent. As described in the main text, we identify the large-scale state by the mean streamwise wave number. The visualizations show that there is also a dependence of the flow field on $\mu$, as can be seen from differences in the streamlines of the flow field as the damping parameter is changed. Because temporal averages may only be computed over finite intervals, the figures do not always exhibit the expected symmetry (e.g.~the different vortices sometimes have different shapes and/or intensities). The length of the temporal averaging interval is determined by the duration for which the individual large-scale structures exist. Because the structures may reform at a shifted streamwise position after a breakdown, averages computed over distinct temporal intervals cannot be merged directly.

The three figures have a tabular structure (with the damping parameter changing over the vertical and the large-scale state over the horizontal).
Missing panels signify that a particular large-scale state has not been clearly observed for a given value of $\mu$ in our data.


\clearpage

\section{Energy budget for various Reynolds numbers}

As explained in the main text, we decompose the generalized Kolmogorov flow velocity
into a $z$-averaged, purely 2D component $\bs v$, the $z$-averaged $z$ component $w$ and the 3D fluctuations $\bs u'$.
The energy distribution among these components, as well as the contributions to the 2D energy budget, Eq.~(3) from the main text, do not vary significantly with the Reynolds number, as can be seen in Supplemental Fig.~\ref{fig:alternative-energetics}.
We show here the contributions to the total energy and the 2D energy budget terms (i.e.~variations of Fig.~3 from the main text) for $Re\approx2850$ and $Re\approx18100$. The results are qualitatively the same.

\begin{figure}[h]
    \begin{tabular}{cc}
        \includegraphics[width =
        0.48\textwidth]{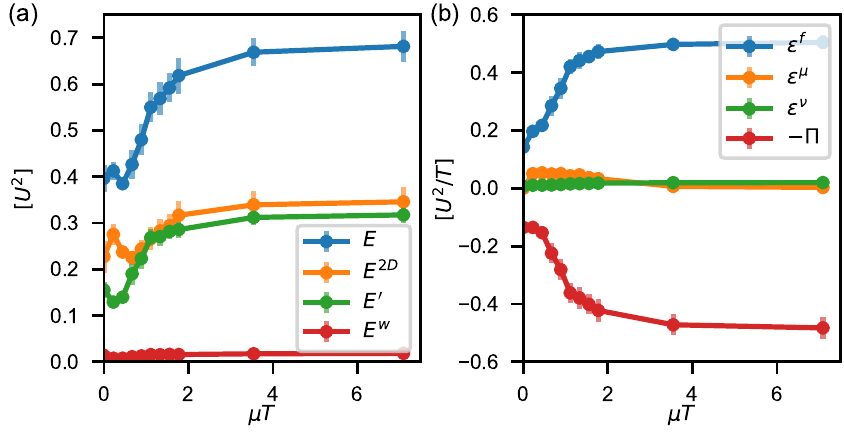}
        &
        \includegraphics[width =
        0.48\textwidth]{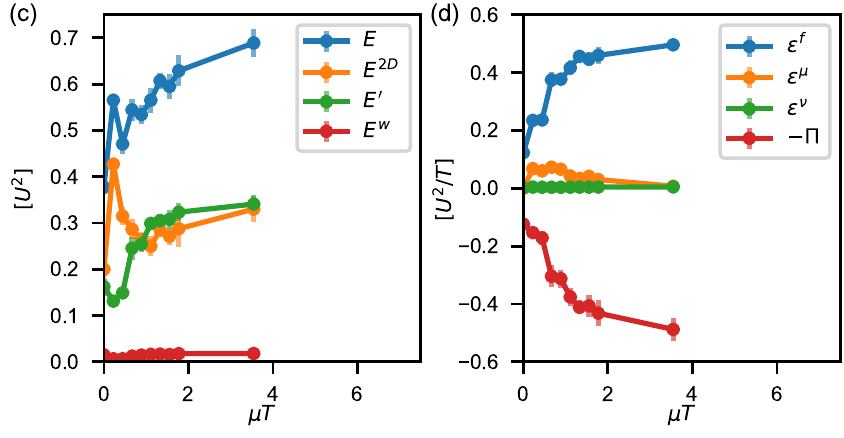}
    \end{tabular}
    \caption{
        Contributions to the total energy and the terms of the 2D energy budget
        (Eq.~(3) in the main text), obtained for $Re\approx2850$ (a,b) and $Re\approx18100$ (c,d), complementing the results for $Re \approx 7180$ in the main text.
        (a, c): The total energy is mostly accounted for by comparable contributions from the 2D flow and the 3D fluctuations.
        The total energy increases as a function of the damping parameter.
        (b, d): Energy budget of the 2D flow. The energy input into the 2D flow increases with increased damping.
        The energy injection rate  $\varepsilon^f$ is almost completely balanced by the transfer of energy from the purely 2D component to the 3D fluctuations, $\Pi$.
        Dissipation through large-scale damping ($\varepsilon^\mu$) and viscous diffusion ($\varepsilon^\nu$) are comparably negligible.
        Error bars span 1 SD computed with respect to temporal variations.
    }
    \label{fig:alternative-energetics}
\end{figure}

\newpage
\section{Vorticity statistics for various Reynolds numbers}
To characterize how the small-scale statistics depend on the damping parameter $\mu$, we present vorticity statistics in Supplemental Fig.~\ref{fig:vorticity_stats_vs_mu}.
As expected, the PDFs become more and more heavy-tailed as the Reynolds number (based on the forcing scales) is increased, in particular the flatness increases with $Re$.
Additionally, we observe a clear trend: At a fixed Reynolds number, the vorticity PDFs exhibit the heaviest tails for small values of $\mu$, and significantly less heavy tails for larger values of $\mu$, despite the fact that the Taylor-scale Reynolds number increases from small non-zero values of $\mu$ to larger values of $\mu$ (cf.~Supplemental Table \ref{tab:simulations}). This is also reflected in the corresponding vorticity flatness. We expect this trend to be the result of the pronounced spatio-temporal inhomogeneity of the large-scale flow. As discussed in the main text, for $\mu T = 0$ the flow is strongly inhomogeneous in space and time due to the emergence and decay of a single vortex pair. For small values of $\mu$, the vortex pair is stabilized, giving rise to a temporally persistent, spatially inhomogeneous large-scale structure. For larger values of $\mu$, two and three vortex pairs emerge (as seen in Supplemental Figs.~\ref{fig:flow_states_N0256}--\ref{fig:flow_states_N1024}), which render the large-scale flow less inhomogeneous.

\begin{figure}[h]
    \begin{tabular}{ccc}
        \includegraphics[width =
        0.325\textwidth]{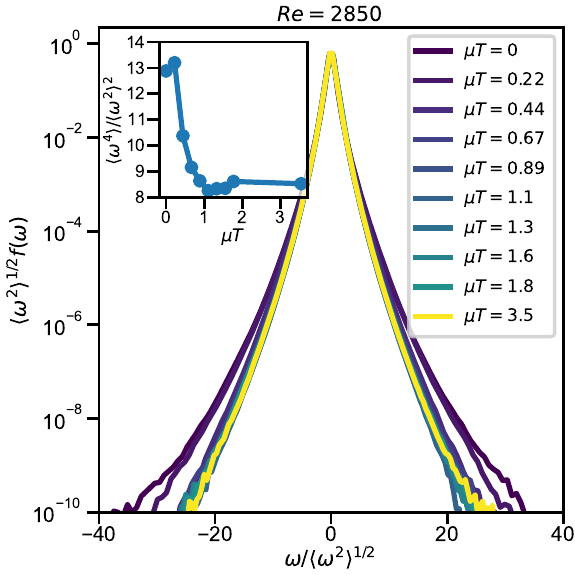}
        &
        \includegraphics[width =
        0.325\textwidth]{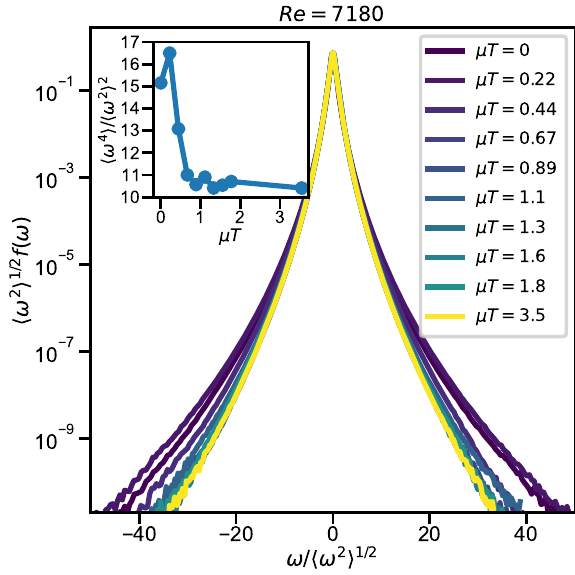}
        &
        \includegraphics[width =
        0.325\textwidth]{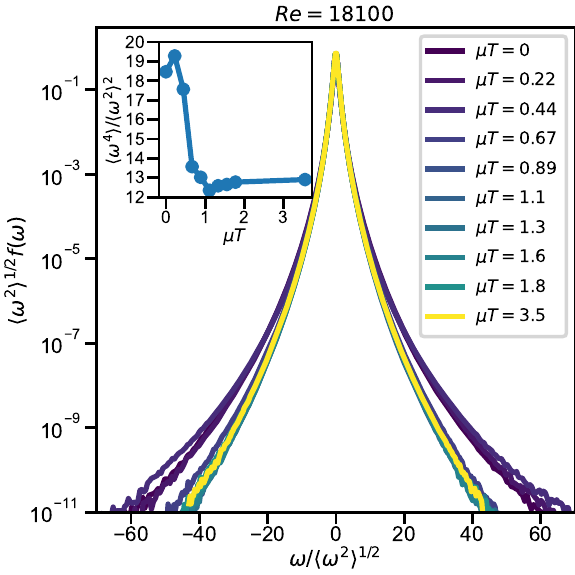}
    \end{tabular}
    \caption{Probability density functions of the vorticity, averaged over space, time and components.
    Insets show corresponding flatness values.}
    \label{fig:vorticity_stats_vs_mu}
\end{figure}

\end{document}